\documentclass{emulateapj}
\usepackage{graphicx} 
\usepackage{color} 
 
\newcommand{\be}{  \begin{eqnarray} }
\newcommand{\ee}{  \end{eqnarray} }
\newcommand{\beq}{\begin{equation}}
\newcommand{\eeq}{\end{equation}}

\def\spose#1{\hbox to 0pt{#1\hss}}
\def\lta{\mathrel{\spose{\lower 3pt\hbox{$\mathchar"218$}}
     \raise 2.0pt\hbox{$\mathchar"13C$}}}
\def\gta{\mathrel{\spose{\lower 3pt\hbox{$\mathchar"218$}}
     \raise 2.0pt\hbox{$\mathchar"13E$}}}

\newcommand{\ltsim}{\lesssim}               
\newcommand{\gtsim}{\gtrsim}                

\newcommand{\achar}{{a_{\rm char}}}
\newcommand{\apeb}{a_p}
\newcommand{\Fp}{{\mathcal F}_p}
\newcommand{\psf}{{\rm psf}}
\newcommand{\xp}{x_p}
\newcommand{\um}{\,\mu{\rm m}}
\newcommand{\mm}{\,{\rm mm}}
\newcommand{\magasecsq}{\,{\rm mag\,arcsec}^{-2}}
\newcommand{\peb}{{\rm p.h.}}
\newcommand{\sky}{{\rm sky}}
\begin{document}

\shorttitle{Brilliant Pebbles}
\title{Brilliant Pebbles: A Method for Detection 
of Very Large Interstellar Grains}
\author{Aristotle Socrates\altaffilmark{1,2} and 
Bruce T. Draine\altaffilmark{2,1}}

\altaffiltext{1}{Institute for Advanced Study, Einstein Drive,
 Princeton, NJ 08540}

\altaffiltext{2}{Department of Astrophysical Sciences, Princeton 
University, Peyton Hall-Ivy Lane, Princeton, NJ 08544; 
socrates@ias.edu; 
draine@astro.princeton.edu}

\begin{abstract}
A photon of wavelength $\lambda\sim 1\um$ 
interacting with a dust grain of radius $a_p\sim 1{\rm mm}$
(a ``pebble'')
undergoes scattering in the forward direction, 
largely within a small characteristic diffraction angle 
$\theta_s\sim \lambda/a_p\sim 100''$.  Though mm-size 
dust grains contribute negligibly to the
interstellar medium's visual 
extinction, the signal they produce in 
scattered light may be detectable, especially for variable 
sources.
Observations of light scattered at
small angles allows for the direct measurement of the 
large grain population; variable sources can also yield
tomographic information of the interstellar medium's mass distribution.    
The ability to detect brilliant pebble halos require a careful 
understanding of the instrument PSF.
 
\end{abstract}

\section{Introduction}

The size distribution of dust in the diffuse interstellar medium
has traditionally been studied through
the wavelength dependence of interstellar extinction and polarization,
and observations of scattered light from reflection nebulae
\citep[cf.][]{Draine_2003a}.
These studies have resulted in models
\citep[][]{Weingartner+Draine_2001a,
               Zubko+Dwek+Arendt_2004,
               Draine+Fraisse_2009}.
where the bulk of the dust mass resides in grains with
radii $a\lesssim 0.5\um$, with approximately 50\% of the grain mass
above and below a characteristic radius $\sim0.15\um$.
To account for the observed extinction, these models tend to consume
the bulk of available elements such as Mg, Si, and Fe in the
``observed'' grain population with $a\ltsim0.5\um$.
Little is known regarding the dust grain size distribution 
above a grain size of order $\sim 0.5\um$, aside from the general 
expectation that grains above this size should contain at most a small 
fraction of the total dust mass.

Therefore, it was surprising when impact detectors on 
{\it Ulysses} and {\it Galileo} measured a flux of
dust particles entering the heliosphere 
\citep{Landgraf+Baggaley+Grun+etal_2000,
       Kruger+Landgraf+Altobelli+Grun_2007}
that 
appeared to indicate that the interstellar grain size distribution 
extended to much larger grains, with approximately
equal mass per unit logarithmic interval out to the largest sizes 
($a\approx 1\um$) that
could be detected.
This finding was completely at odds with the conclusions drawn from studies
of interstellar extinction.
\citet{Draine_2009} argued that the size distribution inferred for the
dust particles approaching the heliosphere could not characterize
average interstellar dust, but the situation remains unclear.
It is further confounded by radar detection of
$a\approx 30\um$ particles 
entering the Earth's atmosphere
on solar-hyperbolic trajectories
\citep{Taylor+Baggaley+Steel_1996,
       Baggaley_2000,
       Baggaley_2004}
implying that they are arriving from interstellar space.
The mass flux in these particles exceeds the mass flux in $0.2 < a<1\um$
particles inferred from the {\it Ulysses} and {\it Galileo} observations.
If these particles are truly entering the solar system from interstellar
space, it implies a total mass density in $a\gtsim 0.5\micron$ grains that,
at least locally, 
exceeds that
in
the $a\ltsim 0.5\um$ grains in conventional models for interstellar grains.
Such an abundance of very large grains would be difficult to understand given
the limitations of interstellar abundances of elements from which such
grains would be constituted.

In this {\it Letter}, we show that very large interstellar grains, 
or ``pebbles,'' are detectable in scattered light,
a phenomenon we refer to
as ``brilliant pebbles.''  The basic idea is presented in
\S\ref{sec:basic}.  Then, in \S\ref{sec:applications} we assess 
the observability of these scattered light halos, and show that detection
is possible even if only a few percent of the interstellar grain mass
is in mm-sized particles.
Variable sources (e.g., gamma-ray bursts ) are particularly useful.
In \S\ref{s: size_dist}
we discuss scattering by
a distribution of pebble sizes.  
We summarize our results in \S\ref{sec:summary}.

\section{Basic Idea}
\label{sec:basic}
Here we show that 
small angle scattering of optical photons by large
dust grains -- ``brilliant pebbles'' -- 
provides a direct method of determining 
the size distribution of very large dust grains, 
and can also provide information on the spatial distribution of the
dust.

Particles with radii $a$ large compared to the wavelength $\lambda$ have
total extinction cross sections equal to twice the geometric area
\citep[the so-called ``extinction paradox'' -- see, e.g.,][]%
{Bohren+Huffman_1983};
 50\%
of the extinction is contributed by diffraction of light, with much of
the energy concentrated in a forward scattering lobe.
The characteristic angular radius for the scattering is
\be
\theta_0 \approx \frac{\lambda}{\pi a} = 
29\arcsec \left(\frac{\lambda}{0.44\micron}\right)
\left(\frac{1\,{\rm mm}}{a}\right) ~~~.
\label{e: basic}
\ee
For small scattering angles $\theta_s$, the differential scattering 
cross section $dC_{\rm sca}/d\Omega \approx constant$,
and for large angles $dC_{\rm sca}/d\Omega \propto \theta_s^{-3}$.
It can be approximated by
\be \label{eq:dCsca/dOmega}
\frac{dC_{\rm sca}}{d\Omega} \approx 
\frac{\pi a^2/\pi \theta_0^2}{1 + (1.8 \theta_s/\theta_0)^3} ~~~.
\ee

\begin{figure}
\epsscale{1.0}
\plotone{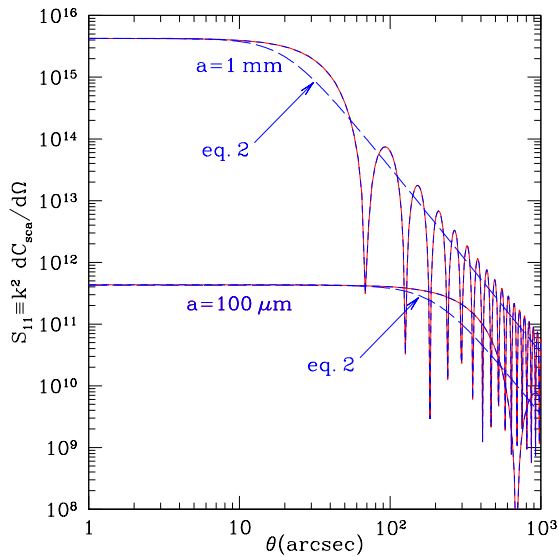}
\caption{Differential scattering cross section vs. scattering angle for 
photons of wavelength
$\lambda=0.55\um$ and spheres
of radii $a=100\um$ and $1$~mm.  The central lobe 
dominates the scattering power in both cases.
The results are insensitive to composition: the figure actually shows
results for both amorphous silicate spheres and graphite spheres, but the
curves fall on top of one another.
Also shown is eq.\ (\ref{eq:dCsca/dOmega}), which is seen to provide
a very good approximation to the overall distribution of scattered power.
\label{fig:halo_scat}}
\end{figure}

Figure \ref{fig:halo_scat} shows exact results for $dC_{\rm sca}/d\Omega$
calculated for spheres using Mie theory
\citep{Mie_1908,
       Bohren+Huffman_1983}.
Results are shown for radii $a=100\um$ and $1$~mm.
Eq.\ (\ref{eq:dCsca/dOmega}) provides a good approximation to
the scattering for single spheres
if one smoothes over the maxima and minima in the diffraction
pattern. 

Assume that ``pebbles'' of radius $\apeb$ account for
a fraction $\Fp$ of the integrated local 
mass density of dust $\rho_d$ where 
\be
\rho_d=\frac{4\pi\rho_0}{3}\int^{\infty}_0 da\, a^3\frac{dn}{da} ~~~,
\\
\Fp\simeq
\frac{\apeb^4 (dn/da)_p}{a^4_{\rm char}(dn/da)_{\rm char}}~~~.
\ee
Here, $\rho_0$ is the density within a grain,
and $\achar\approx 0.15\micron$ 
is the characteristic dust grain radius 
that is responsible for most of the dust mass {\it and} the bulk of the
visual extinction
in the Galaxy's interstellar medium.
Along any given sight line,
the ratio of visual optical depth of pebbles (with radius $a_p$, and number
$\sim a_p (dn/da)_p$) to that of the entire 
grain population is given by
\be
\frac{\Delta\tau_{_V}(\apeb)}{\tau_{_V}}
\simeq \left(\frac{\achar}{\apeb}\right)\Fp \ll 1.
\label{e: depth}
\ee
Therefore such pebbles, if present, contribute negligibly to interstellar
extinction.  

\begin{figure}
\epsscale{1.0}
\plotone{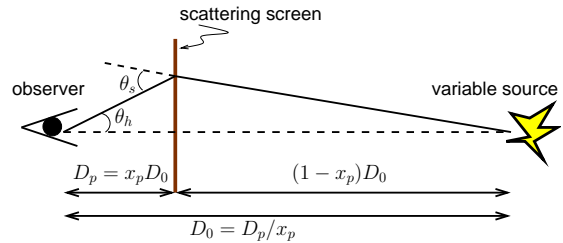}
\caption{Geometry of optical light scattering by
pebbles embedded in a dusty scattering screen.  Examples of the 
variable source are novae, variable stars and GRB optical afterglows.
\label{fig:geometry}}
\end{figure}

Now imagine a point source, that could be variable, as depicted in 
Figure \ref{fig:geometry},  with 
an apparent visual magnitude $m_V$, located at a distance $D_0$
from an observer.  Furthermore, assume that at
distance $D_p$ 
from the observer there is a thin intervening
dusty screen of large angular extent contributing
a visual extinction $A_V$.\footnote{%
      GRB 050724's time-dependent X-ray halo shows 
   that interstellar dust toward this source is distributed in thin sheets
   \citep{Vaughan+Willingale+Romano+etal_2006}.}
Scattering by the pebbles will produce a halo around the source with halo angle
$\theta_h \approx (1-\xp)\theta_s$, where $x_p\equiv D_p/D_0$; 
the characteristic halo angle is
$(1-\xp)\theta_0$.
The time delay $\Delta t_S$ due to small angle 
scattering is
\be
\Delta t_S & \simeq &\frac{1}{1-\xp} \frac{D_p\theta_h^2}{2c}
\\
 & \simeq & \frac{1.2\times 10^4\,{\rm s}}{1-x_p}\,
\left(\frac{D_p}{1\,{\rm kpc}}\right)
\left(\frac{\theta_h}{100\arcsec}\right)^2 ~~~.
\nonumber
\label{e: delta_t}
\ee

With the help of eq. (\ref{e: depth}), 
the photon flux scattered by pebbles of size $\apeb$ is
\be
\label{e: flux}
F_S(t+\Delta t_S) & \simeq & F_0(t)
\left(\frac{\achar}{\apeb}\right)\Fp
\tau_{\rm V}\\
& \simeq & 10^{-5}\,F_0(t)
\left(\frac{\achar/\apeb}{10^{-3}}\right)
\left(\frac{\Fp}{10^{-2}}\right)\left(\frac{\tau_{_{V}}}{1}\right).
\nonumber
\ee   
Note that in the above expression, $F_0(t)$ is the ``direct''   
photon power intercepted by the detector and $F_S(t+\Delta t_S)$ is the 
photon power that underwent small-angle scattering off of pebbles, integrated
over a scattering halo with radius $\theta_h\approx(1-\xp)\theta_s$ with  
surface brightness
\begin{eqnarray}
\nonumber
\frac{\mu_{\lambda,S}(t+\Delta t_S)}{{\rm mag~arcsec}^{-2}}
&\simeq& m_\lambda(t) -
\\
&& 2.5\left[{\rm log}\left( 
\frac{F_S(t+\Delta t_S)({\rm arc~sec})^2}{\pi F_0(t)(1-\xp)^2\theta_0^2}\right)
\right]
\label{e: mu_V}
\end{eqnarray}    
where $m_\lambda(t)$ is the apparent 
magnitude of the point source.
The solid angle $\Delta\Omega_h=\pi\theta_h^2\simeq\pi\theta^2_0(1-\xp)^2$ 
of the scattering halo depends upon the size of the pebbles 
responsible for the small-angle scattering.  With the help of 
eq. (\ref{e: flux}), we note that
\be
\frac{F_S(t+\Delta t_S)}{F_0(t)\pi\theta_h^2} \approx 
\frac{\pi \tau_{\rm V} \,\Fp}{(1-\xp)^2}
\left(\frac{\apeb\achar}{\lambda^2}\right) ~~.
\ee
Somewhat counter-intuitively, for fixed mass fraction 
$\Fp$ the halo surface brightness
{\it increases} with increasing pebble size $\apeb$.  Even though
the total power in scattered light goes as $F_S\propto \apeb^{-1}$, 
the intensity increases $\propto \apeb$ because 
the solid angle that the halo subtends decreases 
as $\apeb^{-2}$.

\section{Astrophysical Applications and Feasibility of Detection}
\label{sec:applications}

In the extreme event that the mass fraction of 
pebbles is 
large such that $\Fp\sim {\mathcal O}(1)$, 
the scattered halo produced by brilliant pebbles is still quite faint.  
By taking an extreme value of $\Fp\sim 1$ in 
eq.\ (\ref{e: flux}), we see that the total power scattered 
is small such that $F_S(t+\Delta t_S)\sim 10^{-3}F_0(t)$.

From Figure {\ref{fig:halo_scat}} we have 
$\theta=(1-\xp)\theta_s$.
Then, for dust in a single screen (i.e., single time delay for given halo angle
$\theta$), the surface brightness of the halo core is
\be \nonumber \label{eq:mu_peb}
\frac{\mu_{\lambda,{\rm max}}(t+\Delta t_s)}{{\rm mag~arcsec}^{-2}} \approx
&& 
\\
\bigg[ m_\lambda(t) + 19.9-2.5\log\Big( &&
\frac{\tau_{\rm V}\Fp}{(1-\xp)^2}
\frac{\apeb}{\mm}
\left(\frac{\um}{\lambda}\right)^2
\Big)\bigg] .~~~
\label{eq:mu_V}
\ee 
Of course, the above expression is independent of time for
a source that does not vary.

\subsection{Strategy and Feasibility for Detection}

How does one detect a faint -- possibly
time variable -- halo?

For $a_p\sim 1\mm$ pebbles, 
the scattering halo is tens of arc seconds in radius.  
%
\citet{Dalcanton+Bernstein_2000} obtained accurate surface photometry
in the {\it B\,-} band -- where the sky is the most dim -- 
of features as faint as 30 mag arcsec$^{-2}$ 
on scales larger than 10$\arcsec$, using exposure times of 45 minutes
on a 2.5m telescope.
At a dark site, the limiting B 
surface brightness (for S/N=1) for a $\sim$10$\arcsec$ region will be
\be\nonumber
\frac{\Delta\mu_{\rm flat}}{{\rm mag~arcsec}^{-2}}&\approx&29.5\,+
2.5\log_{10}(D/2.5{\rm\,m})+
\\
& &1.25\log_{10}(T/45{\rm\,min})
\ee
Thus $\Delta\mu_{\rm flat}\approx 31\,$mag~arcsec$^{-2}$ could be achieved
with a 1 hr exposure on an 8 m telescope.
The limiting factor will not be sky subtraction, but rather subtraction
of the telescope PSF.

The PSF of an optical telescope is
typically characterized by a Gaussian ``core'' produced by 
atmospheric turbulence, a $\theta^{-3}$ halo resulting from diffraction
by the large-scale telescope structure (e.g., aperture)
at intermediate angles, followed by a 
``aureole'', varying roughly as $\theta^{-2}$, due to
a spectrum of small-scale imperfections
(e.g., microripples and dust on the mirrors)
\citep[]{King_1971, Racine_1996, Bernstein_2007}.

The PSF of the Dupont 2.5m telescope at Las Campanas
was measured by
\citet{Bernstein_2007} in 2000 September
($\sim$2 months after mirror realuminization) and 
can be approximately described at large angular 
radii by the following broken power law 
\be \label{eq:psf_inner}
\mu_\psf&\approx& 
m+17+7.5\log_{10}(\theta/40\arcsec) ~~~6\arcsec < \theta < 40\arcsec ~~~
\\
&\approx&  \label{eq:psf_outer}
m+17+5.0\log_{10}(\theta/40\arcsec) ~~~ 
40\arcsec < \theta \ltsim 200\arcsec ~~~.~~~
\ee
where $\mu_\psf$ is in $\magasecsq$.
From here on, we use this exemplary PSF in our estimates.

To estimate exposure times, assume that uncertainties in the sky brightness
on 10$\arcsec$ scales from an exposure time $T_\sky$ is given by
\beq
\frac{\Delta I_\sky}{I_\sky} = C \left(\frac{1}{I_\sky T_\sky}\right)^{1/2}
\eeq
where $C$ is a constant.
Then, on 10$\arcsec$ scales, for an exposure time $T_\psf$,
the uncertainty in the psf intensity at some angle $\theta$ will be given by
\beq
\frac{\Delta I_\psf}{I_\psf+I_\sky} = C 
\left(\frac{1}{(I_\psf+I_\sky) T_\psf}\right)^{1/2}
\eeq
Setting $\Delta I_\psf=I_\peb/(S/N)$, we
can estimate the required exposure time:
\be
T_\psf &=& T_\sky (S/N)^2 
\left(\frac{\Delta I_\sky}{I_\peb}\right)^2
\left(\frac{I_\psf+I_\sky}{I_\sky}\right)
\\
&\approx& 
T_\sky (S/N)^2 
10^{[0.8(\mu_\peb-\Delta\mu_\sky) - 0.4(\mu_\psf-\mu_\sky)]}
\ee
where we assume $I_\psf\gg I_\sky$.
\citet{Dalcanton+Bernstein_2000,Dalcanton+Bernstein_2002} obtained
$\Delta\mu_\sky=29.5\magasecsq$ for $\mu_\sky=22.2\magasecsq$ with
$T_\sky=45~$min on the Dupont 2.5~m.
Thus, if $D$ is the telescope aperture,
\beq \label{eq:exposure}
T_\psf = 0.75 {\rm hr} \left(\frac{2.5\,{\rm m}}{D}\right)^2 (S/N)^2
10^{0.8\mu_\peb-0.4\mu_\psf-14.72}
\eeq
is the exposure time necessary to detect the pebble halo with signal-to-noise
ratio $S/N$, assuming prior knowledge of the telescope psf.

We now assess the detectability of pebble halos.
We set $\apeb=1$~mm, $\Fp=0.1$,
$\lambda=0.44\um$, 
$x_p=0.5$, and $\tau_V=1$ in following estimates.
The characteristic scattering angle $\theta_0=29\arcsec$, resulting
in a scattered halo with nearly constant surface brightness out
to $\theta_h=14.5\arcsec$, with (from eq.\ \ref{eq:mu_peb})
\begin{equation}
\mu_\peb = (m + 19.11)\magasecsq
\end{equation}
For the psf given by eq.\ (\ref{eq:psf_inner},\ref{eq:psf_outer}),
at $\theta=14.5\arcsec$, a steady source has
\beq
\mu_\psf= (m+13.69)\magasecsq .
\eeq

\subsection{Statistical Detection}

The column
density of large dust grains may be correlated 
with the column
density of the sub-$\um$ grains that produce interstellar reddening.
It follows that a statistical search for brilliant pebble halos using large-scale
imaging, such as the Sloan Digital Sky Survey, may be fruitful.  
Rather than fit for only
an ``intrinsic'' PSF $P_0(\theta,\phi)$
that is the same for every star in a field 
(or at most dependent only on position within the field of view), one could
determine whether the actual images (excluding regions with small-scale nebular
emission) can be better reproduced using a stellar
PSF $P_0(\theta,\phi)+E(B-V)P_1(\theta)$, 
where 
\begin{equation}
P_1(\theta_h)= \frac{1}{E(B-V)}\int_0^1 dx_p \frac{f(x_p)}{(1-x_p)^2}
\left(\frac{d\tau_{\rm sca}}{d\Omega}\right)_{\theta_s}
\end{equation}
is the dust contribution to the PSF,
per unit $E(B-V)$, where
$f(x)dx$ is the fraction of the reddening contributed by
dust with $x_p$ in $[x,x+dx]$.
For a first search, it would be adequate to assume the dust to be
halfway to the star (i.e., $f(x)\rightarrow\delta(x-0.5)$).
Using a large number of stars to determine $P_1(\theta)$
may allow detection of the ``brilliant pebble'' phenomenon at levels
that might be impossible for single stars.

\subsection{Bright Stars}
The pebble halo can be detected with reasonable exposure times on individual
bright stars.  For a star with $A_V\approx 1\,$mag and $m_B < 8.5\,$mag 
(so that $\mu_\psf < \mu_\sky$)
eq.\ (\ref{eq:exposure}) becomes
\begin{equation}
\left(\frac{T}{\rm hr}\right) 
\approx
0.75\,{\rm hr}\left(S/N\right)^2
\left(\frac{2.5{\rm \, m}}{D}\right)^2
10^{0.4m_B - 4.91}
\end{equation}
Thus an exposure time of only 1.4 minutes would suffice for detection
(with $S/N=5$) of
the pebble halo around a star with $m_B=5\,$mag.  
The challenge will be to
know the telescope psf well enough to accurately subtract it.  This can be
accomplished by measuring the psf using stars with minimal reddening.
Because the uncertainty in the psf should be small compared to
$I_\peb/I_\psf\approx 0.007 (A_V/{\rm mag})$,
determination of the psf will require care.

\subsection{GRB Optical Afterglows and Other Bright Transients}

If the source is variable, the ability to detect 
a brilliant pebble halo of scattered light may increase dramatically.
If the PSF of the telescope and atmosphere is stable and known,
then point sources whose angular positions
are fixed can be subtracted from each successive exposure.  In doing so, 
the pebble halo, which varies in both time and space, may be isolated 
from the persistent stellar and interstellar background light.

Gamma-Ray bursts (GRBs) occur at a rate $\sim$ once per day in the 
Universe.  
Some GRB
afterglows have optical luminosities that 
can compete with those of quasars.  For example GRB 990123
reached a peak optical flux slightly brighter than $m_V=9\,$mag
\citep{Akerlof+Balsano+Barthelmy+etal_1999},
and GRB 080319B had $m_V < 6\,$mag for $\sim 40\sec$
\citep{Bloom+Perley+Li+etal_2008}.
There may also be other very bright optical transients:
e.g.,
\citet{Shamir+Nemiroff_2006} observed what appeared to be a 5th magnitude
flash that lasted $\sim10\,$min.
In the event that a burst with $m_B=5\,$mag were to 
occur at low Galactic latitudes, the rapid fading of the point source
provides the ideal
circumstance for detection of a pebble halo, 
because observation of the time-delayed
scattered-light halo can be done without interference from the telescope psf.
If an optical transient shone with $m_B=5$ for a few minutes and then faded.
the time-delayed scattered light ``halo'' (with $\mu=24.1\magasecsq$ if dust
with $A_V=1$ is present in a thin sheet) could
be detected in $\sim$2~min of integration on an 8~m class telescope.

\subsection {Variable Stars}
 
There are several thousand variable stars in the Milky Way
brighter than $m_V=10$ 
\citep{Paz_2006}. 
In fact, from the {\it ASAS} catalogue, there are 
over 400 eclipsing binaries with periods shorter than 
1 day and $V<10$.  For $\sim100\micron$ pebbles
the delay timescale is in terms of hours, rather than 
tens of minutes.  It therefore follows that 
the pebble halo can be detected by image subtraction
if the binary is significantly reddened and distant, 
if $\sim100\micron$ grains are
sufficiently numerous.

\section{Small angle scattering from a size distribution of pebbles}
\label{s: size_dist}

So far, we have considered small angle scattering 
due to the presence of pebbles of a single characteristic 
size.  Now, we briefly consider a
distribution of sizes.  For simplicity, we assume 
that the pebble size distribution resembles a power law
\be
\frac{dN}{da}= {\mathcal N} a^{-p}   ~~~~a_{\rm min} < a < a_{\rm max}
\ee 
where $N(a)$ is the column density of pebbles with radii $\apeb<a$.
For $p=4$ (i.e., equal mass per logarithmic interval in $\apeb$)
the normalization constant  
is given by 
\be
{\mathcal N} \approx \frac{\tau_V}{2\pi}\frac{\achar}
{\ln(a_{\rm max}/a_{\rm min})}
\Fp 
\label{e: normal}
\ee
where the optical depth $\tau_V$ is assumed to be provided by
the dominant dust population with radii $a_{\rm char}\approx 0.15\micron$,
with extinction cross sections $C_{\rm ext}\approx \pi a_{\rm char}^2$.

For scattering by material in one scattering screen, 
we define the differential scattering optical depth such that, at halo
angle $\theta_h$, the scattered intensity is
\be
I(\theta_h,t+\Delta t_s) = \frac{F_0(t)}{(1-x_p)^2}
\left(\frac{d\tau_{\rm sca}}{d\Omega}\right)_{\theta_h}
\ee
where 
\be
\left(\frac{d\tau_{\rm sca}}{d\Omega}\right)_{\theta_h}
=
\int da \left(\frac{dN}{da}\right) 
\left(\frac{dC_{\rm sca}}{d\Omega}\right)_{\theta_s=\theta_h/(1-x_p)}
\ee
With the help of the approximation (\ref{eq:dCsca/dOmega})
for $dC_{\rm sca}/d\Omega$, the
differential scattering optical depth is
\be
\label{e: dCscatt_dist}
\frac{d\tau_{\rm sca}}{d\Omega}&\simeq&
\mathcal{N} \frac{\pi^2}{\lambda^2} 
\int_{a_{\rm min}}^{a_{\rm max}} \frac{a^{4-p} da}{1+(a/a_c)^3}
\\
a_c&\equiv&\frac{(1-\xp)\lambda}{1.8\pi\theta_h} ~~~.
\ee
We can distinguish three regimes: the core, with 
$\theta_h < \theta_1 \equiv (1-\xp)\lambda/1.8\pi a_{\rm max}$;
the intermediate halo, with 
$\theta_1 < \theta_h < \theta_2 \equiv (1-\xp)\lambda/1.8\pi a_{\rm min}$;
and the outer halo, with $\theta_h > \theta_2$.
For $2 < p < 5$ we have
\be
\frac{d\tau_{\rm sca}}{d\Omega} &\approx & 
\mathcal{N}\frac{\pi^2 a_{\rm max}}{\lambda^2}
 ~~~{\rm for~} \theta_h \ll \theta_1
\\ \nonumber
&\approx& 
\mathcal{N}\frac{2\pi^2(1-\xp)^{5-p}}{(5-p)(p-2)(1.8\pi)^{5-p}}
\frac{1}{\lambda^{p-3}}\frac{1}{\theta_h^{5-p}}
\\
&&~~~~~~~~~~~~~~~~~~~~{\rm for~} \theta_1 \ll \theta_h \ll \theta_2
\\
&\approx& 
\mathcal{N}\frac{\pi^2 (1-\xp)^3}{(p-2)(1.8\pi)^3a_{\rm min}^{p-2}}
\frac{\lambda}{\theta_h^3}
~~~{\rm for~} \theta_h \gg \theta_2
\ee

\begin{figure}[t]
\epsscale{1.00}
\plotone{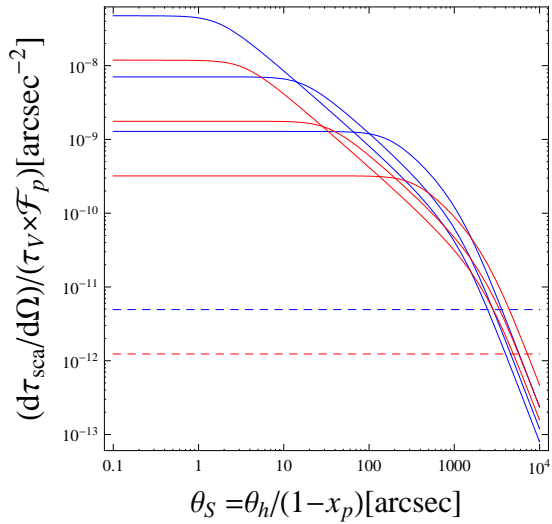}
\caption{Pebble size-averaged 
differential optical depth $(d\tau_{\rm sca}/d\Omega)/(\tau_V\mathcal{F}_p)$
vs.\ scattering angle $\theta_s$ for 
photons of wavelength $\lambda=0.4405\um$ (blue) and $\lambda=
0.802\um$ (red).  
Each curve represents $dC_{\rm sca}/d\Omega$, integrated
over an $a^{-4}$ pebble size distribution, but with varying 
ranges in pebble size.  The lower limit for pebble sizes $a_{\rm min}
=10\um$ for all of the curves.  The maximum pebble 
size $a_{\rm max}=1\,{\rm cm}$, $1\,{\rm mm}$ 
and $10^2\um$ for the
top, middle and bottom curves, respectively, for each of the two 
photon wavelengths.  At small scattering angles, 
$d\tau_{\rm sca}/d\Omega\propto$ a constant, while 
$d\tau_{\rm sca}/d\Omega\propto \theta^{-3}_s$ at large scattering 
angles, recovering the asymptotic behavior of eq. (\ref{eq:dCsca/dOmega}).
At intermediate angles, $d\tau_{\rm sca}/d\Omega\propto
\theta^{-1}_s$.  This `intermediate'' angle range is more prominent 
for broad distributions of pebble sizes.
The dashed lines at the bottom represent the limit where the 
pebbles are replaced purely by ``normal'' $a\simeq 0.15\um$ grains for 
each photon band.  As long as ${\mathcal F}_p\gtrsim 10^{-2}$ for
the pebble-size distributions utilized in this figure, then 
the variable scattered light should be dominated by pebbles
rather than normal dust.  }
\label{f: size_dist}. 
\end{figure}

In Figure \ref{f: size_dist}, we perform the integral 
in eq. (\ref{e: dCscatt_dist}) over a $p=4$ pebble size 
distribution for 
$\lambda=0.4405\um$ and $0.802\um$.  
As an illustrative example, 
consider the case in Figure \ref{f: size_dist} 
where the large pebble size cutoff $a_{\rm max}=1.0$ cm
for $\lambda=0.4405\um$.  
The angle $\theta_1$ marking the transition from the
the core,
where $d\tau_s/d\Omega\approx const$,
to the intermediate halo, where $d\tau_s/d\Omega \propto \theta_h^{5-p}$,
indicates the value of the 
upper cutoff in pebble size, $a_{\max}\approx (1-\xp)\lambda/1.8\pi\theta_1$.

The power-law index $p$ for the size distribution can be determined
from both the angular dependence $\theta_h^{p-5}$ and
the color ($d\tau_s/d\Omega\propto \lambda^{3-p}$)
of the intermediate halo ($\theta_1 \ll \theta_h \ll \theta_2$).
The large angle behavior $d\tau_{\rm sca}/d\Omega\propto\theta^{-3}$
for $\theta\gtrsim\theta_2$
indicates that at these angles, all of the pebble size range
$[a_{\rm min},a_{\rm max}]$ is in
the large scattering angle regime.

Figure \ref{f: size_dist} shows how
$d\tau_{\rm sca}/d\Omega$ varies with photon
wavelength.
For shorter wavelengths, 
the halo makes the transition from the small angle to intermediate
angle regime at smaller angles.
The scattered halo will be very blue in the core:
for $p=4$ the core will have a color excess
$B-I=-2.5\log_{10}(0.8020/.4405)^2=-1.30\,$mag relative to the
unscattered source, while the intermediate zone beyond the core
will have $B-I=-0.65\,$mag.
For $\theta > \theta_2$ the halo will be red, with $B-I=+0.65\,$mag,
but the
combination of long delay time $\Delta t_S$ and low surface brightness may
render this outer halo undetectable.

\section{Summary}
\label{sec:summary}

We propose a method for the detection of extremely 
large ($\sim 0.01-1$ cm) dust grains.
When the ratio of photon wavelength to grain size $\lambda/a\ll 1$,
scattering of light is highly peaked in the forward direction, with 
a characteristic scattering angle 
$\theta_0\sim\lambda/\pi\,a\sim 30(1\,{\rm mm}/\apeb)\arcsec$.
For a variable source, the
halo will lag with a
characteristic time-scale $\sim 10$ min for scattering by 1~mm pebbles
at a distance $\sim 1$ kpc.  We refer to 
this phenomena as ``brilliant pebbles.''

Detection of the brilliant pebble halo will be technically
challenging.  Even for favorable assumptions regarding the pebble size
distribution, an accurate determination of the telescope PSF is
required in order to detect such a low surface brightness signal.
Nevertheless, we show that a pebble halo could be detected using
existing facilities if the mass in $\sim$1~mm pebbles is more than a
few percent of the total dust mass, as has been suggested by
observations of high-velocity dust grains in interplanetary space and
hyperbolic micrometeors (see \S~1).

Detection is best accomplished using a well-baffled 
telescope with a freshly aluminized
mirror, to minimize scattering by small-scale imperfections such as
dust on the mirror.

A brief but bright optical burst would be ideal, as the direct light
can fade while the time-delayed pebble halo persists, allowing the
telescope PSF and the pebble halo to be separated in the time domain.
Also, the pebble halo from a bright short period eclipsing eclipsing binary 
that is significantly reddened can be subtracted from the PSF in the time 
domain as well.  

If a ``brilliant pebble'' halo is 
detected, the angular variation of the halo intensity
will indicate the pebble size range and the form of the particle size
distribution.  The halo will be very blue in the core, and moderately blue
in the intermediate halo; the color of the intermediate halo provides
an independent measure of the size distribution.
If the source is suitably variable, 
the distance to the pebbles can be determined from
the time delay of the halo relative to the point source.

Upper limits on brilliant pebble halos
can provide valuable constraints
on the size distribution of solid particles in the interstellar medium.

\acknowledgements{We thank P. Chang, M. Pan and A. Witt 
for fruitful discussions and the referee, whose comments improved
our paper.  AS acknowledges support from a Lyman 
Spitzer Jr. Fellowship given by Astrophysical Sciences at Princeton 
U. and a Friends of the Institute 
Fellowship at IAS, Princeton.
BTD thanks the IAS for hospitality during summer 2008 and 
support from NSF grant AST-0406883.}


\begin{thebibliography}{22}
\expandafter\ifx\csname natexlab\endcsname\relax\def\natexlab#1{#1}\fi

\bibitem[{{Akerlof} {et~al.}(1999){Akerlof}, {Balsano}, {Barthelmy}, {Bloch},
  {Butterworth}, {Casperson}, {Cline}, {Fletcher}, {Frontera}, {Gisler},
  {Heise}, {Hills}, {Kehoe}, {Lee}, {Marshall}, {McKay}, {Miller}, {Piro},
  {Priedhorsky}, {Szymanski}, \& {Wren}}]{Akerlof+Balsano+Barthelmy+etal_1999}
{Akerlof}, C., {et~al.} 1999, \nat, 398, 400

\bibitem[{{Baggaley}(2000)}]{Baggaley_2000}
{Baggaley}, W.~J. 2000, \jgr, 105, 10353

\bibitem[{{Baggaley}(2004)}]{Baggaley_2004}
---. 2004, Earth Moon and Planets, 95, 197

\bibitem[{{Bernstein}(2007)}]{Bernstein_2007}
{Bernstein}, R.~A. 2007, \apj, 666, 663

\bibitem[{{Bloom} {et~al.}(2008){Bloom}, {Perley}, {Li}, {Butler}, {Miller},
  {Kocevski}, {Kann}, {Foley}, {Chen}, {Filippenko}, {Starr}, {Macomber},
  {Prochaska}, {Chornock}, {Poznanski}, {Klose}, {Skrutskie}, {Lopez}, {Hall},
  {Glazebrook}, \& {Blake}}]{Bloom+Perley+Li+etal_2008}
{Bloom}, J.~S., {et~al.} 2008, ArXiv e-prints

\bibitem[{Bohren \& Huffman(1983)}]{Bohren+Huffman_1983}
Bohren, C.~F., \& Huffman, D.~R. 1983, Absorption and Scattering of Light by
  Small Particles (New York: Wiley)

\bibitem[{{Dalcanton} \& {Bernstein}(2000)}]{Dalcanton+Bernstein_2000}
{Dalcanton}, J.~J., \& {Bernstein}, R.~A. 2000, \aj, 120, 203

\bibitem[{{Dalcanton} \& {Bernstein}(2002)}]{Dalcanton+Bernstein_2002}
---. 2002, \aj, 124, 1328

\bibitem[{{Draine}(2003)}]{Draine_2003a}
{Draine}, B.~T. 2003, \araa, 41, 241

\bibitem[{{Draine}(2009)}]{Draine_2009}
---. 2009, Space Science Reviews, 143, 333

\bibitem[{{Draine} \& {Fraisse}(2009)}]{Draine+Fraisse_2009}
{Draine}, B.~T., \& {Fraisse}, A.~A. 2009, \apj, 696, 1

\bibitem[{{King}(1971)}]{King_1971}
{King}, I.~R. 1971, \pasp, 83, 199

\bibitem[{{Kr{\"u}ger} {et~al.}(2007){Kr{\"u}ger}, {Landgraf}, {Altobelli}, \&
  {Gr{\"u}n}}]{Kruger+Landgraf+Altobelli+Grun_2007}
{Kr{\"u}ger}, H., {Landgraf}, M., {Altobelli}, N., \& {Gr{\"u}n}, E. 2007,
  Space Science Reviews, 130, 401

\bibitem[{{Landgraf} {et~al.}(2000){Landgraf}, {Baggaley}, {Gr{\"u}n},
  {Kr{\"u}ger}, \& {Linkert}}]{Landgraf+Baggaley+Grun+etal_2000}
{Landgraf}, M., {Baggaley}, W.~J., {Gr{\"u}n}, E., {Kr{\"u}ger}, H., \&
  {Linkert}, G. 2000, \jgr, 105, 10343

\bibitem[{{Mie}(1908)}]{Mie_1908}
{Mie}, G. 1908, Annalen der Physik, 330, 377

\bibitem[{{Paczy{\'n}ski} {et~al.}(2006){Paczy{\'n}ski}, {Szczygie{\l}},
  {Pilecki}, \& {Pojma{\'n}ski}}]{Paz_2006}
{Paczy{\'n}ski}, B., {Szczygie{\l}}, D.~M., {Pilecki}, B., \& {Pojma{\'n}ski},
  G. 2006, \mnras, 368, 1311

\bibitem[{{Racine}(1996)}]{Racine_1996}
{Racine}, R. 1996, \pasp, 108, 699

\bibitem[{{Shamir} \& {Nemiroff}(2006)}]{Shamir+Nemiroff_2006}
{Shamir}, L., \& {Nemiroff}, R.~J. 2006, \pasp, 118, 1180

\bibitem[{{Taylor} {et~al.}(1996){Taylor}, {Baggaley}, \&
  {Steel}}]{Taylor+Baggaley+Steel_1996}
{Taylor}, A.~D., {Baggaley}, W.~J., \& {Steel}, D.~I. 1996, \nat, 380, 323

\bibitem[{{Vaughan} {et~al.}(2006){Vaughan}, {Willingale}, {Romano}, {Osborne},
  {Goad}, {Beardmore}, {Burrows}, {Campana}, {Chincarini}, {Covino}, {Moretti},
  {O'Brien}, {Page}, {Supper}, \&
  {Tagliaferri}}]{Vaughan+Willingale+Romano+etal_2006}
{Vaughan}, S., {et~al.} 2006, \apj, 639, 323

\bibitem[{{Weingartner} \& {Draine}(2001)}]{Weingartner+Draine_2001a}
{Weingartner}, J.~C., \& {Draine}, B.~T. 2001, \apj, 548, 296

\bibitem[{{Zubko} {et~al.}(2004){Zubko}, {Dwek}, \&
  {Arendt}}]{Zubko+Dwek+Arendt_2004}
{Zubko}, V., {Dwek}, E., \& {Arendt}, R.~G. 2004, \apjs, 152, 211

\end{thebibliography}

\end{document}